\documentclass{paperClass}

\begin{document}
\newcommand{\berlin}{Institut f\"{u}r Physik, Humboldt-Universit\"{a}t zu Berlin, Newtonstr.~~15, 12489 Berlin, Germany}
\newcommand{\bonn}{Helmholtz-Institut f\"{u}r Strahlen- und Kernphysik, Rheinische Friedrich-Wilhelms-Universit\"{a}t Bonn, Nussallee 14-16, 53115 Bonn, Germany}
\newcommand{\cmu}{Department of Physics, Carnegie Mellon University, Pittsburgh, PA 15213, USA}
\newcommand{\cwru}{Department of Physics, Case Western Reserve University, Cleveland, OH 44106, USA}
\newcommand{\etp}{Institute of Experimental Particle Physics~(ETP), Karlsruhe Institute of Technology~(KIT), Wolfgang-Gaede-Str.~1, 76131 Karlsruhe, Germany}
\newcommand{\fulda}{University of Applied Sciences~(HFD)~Fulda, Leipziger Str.~123, 36037 Fulda, Germany}
%
%%% BEGIN: KIT institutes
\newcommand{\iap}{Institute for Astroparticle Physics~(IAP), Karlsruhe Institute of Technology~(KIT), Hermann-von-Helmholtz-Platz 1, 76344 Eggenstein-Leopoldshafen, Germany}
\newcommand{\ipe}{Institute for Data Processing and Electronics~(IPE), Karlsruhe Institute of Technology~(KIT), Hermann-von-Helmholtz-Platz 1, 76344 Eggenstein-Leopoldshafen, Germany}
\newcommand{\itep}{Institute for Technical Physics~(ITEP), Karlsruhe Institute of Technology~(KIT), Hermann-von-Helmholtz-Platz 1, 76344 Eggenstein-Leopoldshafen, Germany}
\newcommand{\ppq}{Project, Process, and Quality Management~(PPQ), Karlsruhe Institute of Technology~(KIT), Hermann-von-Helmholtz-Platz 1, 76344 Eggenstein-Leopoldshafen, Germany    }
%
%%% END: KIT Institutes
%
\newcommand{\inr}{Institute for Nuclear Research of Russian Academy of Sciences, 60th October Anniversary Prospect 7a, 117312 Moscow, Russia}
\newcommand{\lbnl}{Institute for Nuclear and Particle Astrophysics and Nuclear Science Division, Lawrence Berkeley National Laboratory, Berkeley, CA 94720, USA}
\newcommand{\madrid}{Departamento de Qu\'{i}mica F\'{i}sica Aplicada, Universidad Autonoma de Madrid, Campus de Cantoblanco, 28049 Madrid, Spain}
\newcommand{\mainz}{Institut f\"{u}r Physik, Johannes-Gutenberg-Universit\"{a}t Mainz, 55099 Mainz, Germany}
\newcommand{\mpp}{Max-Planck-Institut f\"{u}r Physik, F\"{o}hringer Ring 6, 80805 M\"{u}nchen, Germany}
\newcommand{\massit}{Laboratory for Nuclear Science, Massachusetts Institute of Technology, 77 Massachusetts Ave, Cambridge, MA 02139, USA}
\newcommand{\mpik}{Max-Planck-Institut f\"{u}r Kernphysik, Saupfercheckweg 1, 69117 Heidelberg, Germany}
\newcommand{\muenster}{Institute for Nuclear Physics, University of M\"{u}nster, Wilhelm-Klemm-Str.~9, 48149 M\"{u}nster, Germany}
\newcommand{\npi}{Nuclear Physics Institute,  Czech Academy of Sciences, 25068 \v{R}e\v{z}, Czech Republic}
\newcommand{\unc}{Department of Physics and Astronomy, University of North Carolina, Chapel Hill, NC 27599, USA}
\newcommand{\washington}{Center for Experimental Nuclear Physics and Astrophysics, and Dept.~of Physics, University of Washington, Seattle, WA 98195, USA}
\newcommand{\wuppertal}{Department of Physics, Faculty of Mathematics and Natural Sciences, University of Wuppertal, Gau{\ss}str.~20, 42119 Wuppertal, Germany}
\newcommand{\saclay}{IRFU (DPhP \& APC), CEA, Universit\'{e} Paris-Saclay, 91191 Gif-sur-Yvette, France }
\newcommand{\tum}{Technische Universit\"{a}t M\"{u}nchen, James-Franck-Str.~1, 85748 Garching, Germany}
\newcommand{\lmu}{Ludwig-Maximilians-Universit\"{a}t M\"{u}nchen, Geschwister-Scholl-Platz 1, 80539 M\"{u}nchen, Germany}
\newcommand{\uhd}{Institute for Theoretical Astrophysics, University of Heidelberg, Albert-Ueberle-Str.~2, 69120 Heidelberg, Germany}
\newcommand{\tunl}{Triangle Universities Nuclear Laboratory, Durham, NC 27708, USA}
%
%%% BEGIN: other institutions
%
\newcommand{\ornl}{Also affiliated with Oak Ridge National Laboratory, Oak Ridge, TN 37831, USA}
%
%\newcommand{\swansea}{Department of Physics, Swansea University, Singleton Park, Swansea SA2 8PP, United Kingdom}
%\newcommand{\ucsb}{Department of Physics, University of California at Santa Barbara, Santa Barbara, CA 93106, USA}
%
%%% END: other institutions
%

%\affiliation{\tlk}
\affiliation{\iap}
\affiliation{\ipe}
\affiliation{\inr}
\affiliation{\muenster}
\affiliation{\etp}
\affiliation{\itep}
\affiliation{\tum}
\affiliation{\mpp}
\affiliation{\unc}
\affiliation{\tunl}
\affiliation{\lbnl}
\affiliation{\wuppertal}
\affiliation{\madrid}
\affiliation{\washington}
\affiliation{\npi}
\affiliation{\massit}
\affiliation{\cmu}
\affiliation{\saclay}
\affiliation{\mpik}
\affiliation{\berlin}
\affiliation{\uhd}
\affiliation{\mainz}

%\affiliation{\fulda}
%\affiliation{\bonn}
%\affiliation{\cwru}
%\affiliation{\ppq}

\author{M.~Aker}\affiliation{\iap}
\author{D.~Batzler}\affiliation{\iap}
\author{A.~Beglarian}\affiliation{\ipe}
\author{J.~Behrens}\affiliation{\iap}
\author{A.~Berlev}\affiliation{\inr}
\author{U.~Besserer}\affiliation{\iap}
\author{B.~Bieringer}\affiliation{\muenster}
\author{F.~Block}\affiliation{\etp}
\author{S.~Bobien}\affiliation{\itep}
\author{B.~Bornschein}\affiliation{\iap}
% ...

\collaboration{KATRIN Collaboration}\noaffiliation

\title{Search for Lorentz-Invariance Violation with the first KATRIN data}

\begin{abstract}
%Some quantum gravity theories predict deviations from CPT and Lorentz symmetry \cite{PhysRevD.39.683,PhysRevD.51.3923,Carroll_2001,Gambini_1999}. In the neutrino sector strong constraints have been set by time-of-flight and oscillation experiments. In this work we address the parameters $\left(a_{\text {of}}^{(3)}\right)_{00}$, $\left(a_{\text {of}}^{(3)}\right)_{10}$, and $\left(a_{\text {of}}^{(3)}\right)_{11}$ which are not accessible in these experiments \cite{Diaz}. They could however manifest themselves in a non-isotropic beta-decaying source as a sidereal oscillation of a the spectral endpoint. From the data of the first scientific run of the Karlsruhe Neutrino Experiment KATRIN, the parameter $\abs{\left(a_{\text{of}}^{(3)}\right)_{11}}$ is limited to $\lesssim \SI{3.7e-6}{\giga \electronvolt}$. This is the first limit on this parameter.

Some extensions of the Standard Model of Particle Physics allow for Lorentz invariance and Charge-Parity-Time (CPT)-invariance violations. In the neutrino sector strong constraints have been set by neutrino-oscillation and time-of-flight experiments. However, some Lorentz-invariance-violating parameters are not accessible via these probes. In this work, we focus on the parameters $(a_{\text{of}}^{(3)})_{00}$, $(a_{\text{of}}^{(3)})_{10}$ and $(a_{\text{of}}^{(3)})_{11}$ which would manifest themselves in a non-isotropic $\upbeta$-decaying source as a sidereal oscillation and an overall shift of the spectral endpoint. Based on the data of the first scientific run of the KATRIN experiment, we set the first limit on $\left|(a_{\text{of}}^{(3)})_{11}\right|$ of $< \SI{3.7e-6}{\giga \electronvolt}$ at 90\% confidence level. Moreover, we derive new constraints on $(a_{\text{of}}^{(3)})_{00}$ and $(a_{\text{of}}^{(3)})_{10}$.
\end{abstract}

\maketitle

\section{Introduction}
CPT and Lorentz invariance are central ingredients of modern physics and of the Standard Model (SM) of particle physics. However, some extensions of the SM such as string theories \cite{PhysRevD.39.683,PhysRevD.51.3923}, loop quantum gravity \cite{Gambini_1999}, and non-commutative quantum field theories~\cite{Carroll_2001} suggest that Charge-Parity-Time (CPT) and Lorentz invariance may be violated at high energies. Yet, so far, no experimental evidence for CPT or Lorentz invariance violation was ever observed and the parameter space is strongly constrained. 
%limits on the parameter space were set.
%Also from a theoretical point of view one expects that in effective field theories these Lorentz invariance violating effects are heavily suppressed by a factor which connects the electroweak and Planck scales $m_W/m_P \approx 10^{-17}$~\cite{PhysRevD.51.3923}.

Deviations from Lorentz symmetry are typically described in a relativistic effective field theory, the so-called Standard Model Extension (SME)~\cite{PhysRevD.55.6760,PhysRevD.58.116002,PhysRevD.69.105009}. In particular, the SME specifies all possible Lorentz-invariance-violating operators for neutrino propagation, many of which have been constrained with neutrino-oscillation experiments \cite{PhysRevD.85.096005}. The so-called  \enquote{oscillation-free} modes, which cannot be assessed via oscillation experiments, are usually constrained by time-of-flight experiments, which probe the neutrinos' group velocity compared to that of photons. However, there are four oscillation-free parameters, $(a_{\text {of}}^{(d=3)})_{jm}$, where $j$ and $m$ denote the angular momentum quantum numbers with $j=0,1$ and d stands for the mass dimension. These parameters can only be accessed by interaction processes, such as the $\beta$-decay of tritium~\cite{Diaz,Lehnert:2021tbv}. 

\begin{figure}[]
\centering
\includegraphics[width=.4\textwidth]{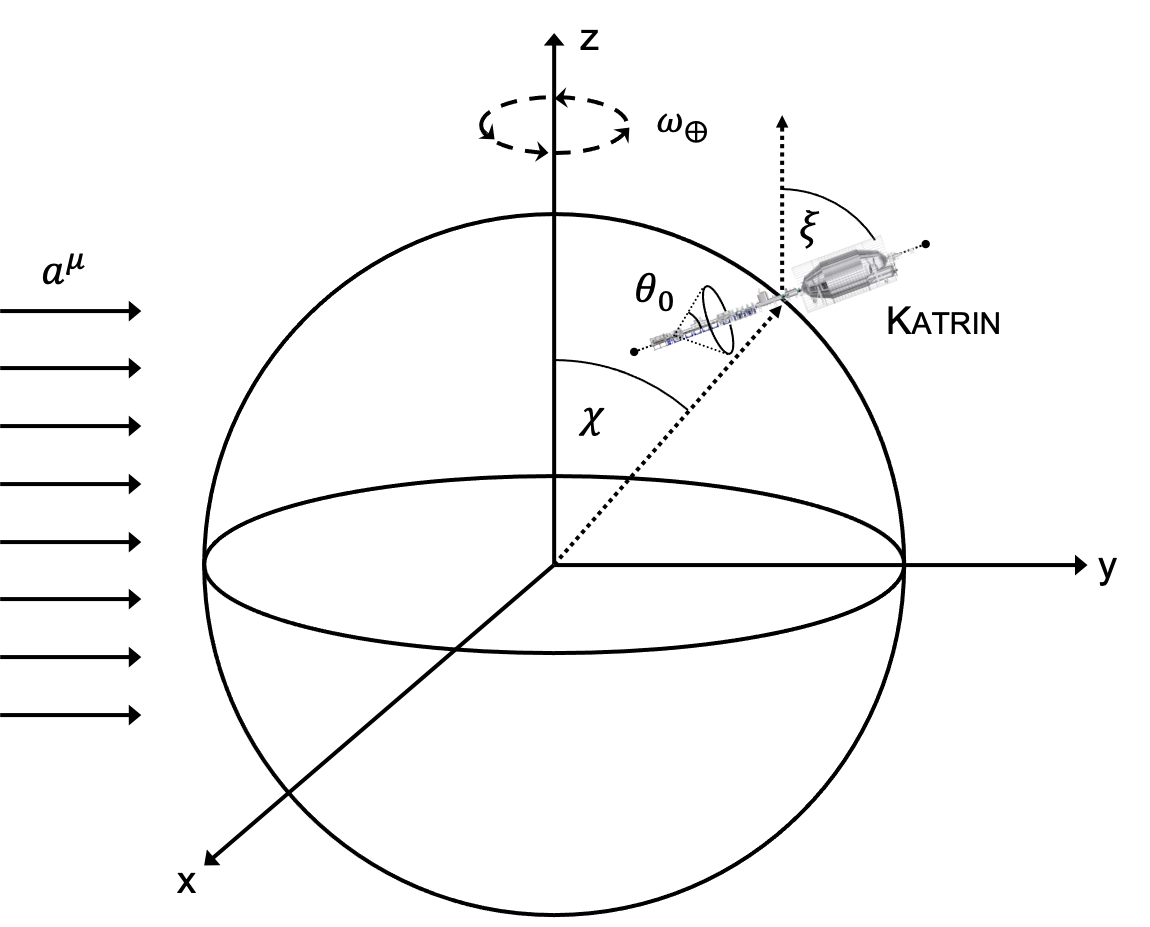}
\caption{Sketch of the equatorial coordinate system. The Earth is rotating with $\omega_{\oplus}$ within the Lorentz-invariance-violating vector field $a^{\mu}$, defined to be perpendicular to the z-axis. Therefore the KATRIN experiment, which is located at the colatitude $\chi \approx \SI{41}{\degree}$, is moving with velocity $\beta_{\text{rot}}$ (rotation velocity of the Earth at the position of KATRIN). The beam-axis of the KATRIN experiment is tilted with respect to the local north by $\xi \approx \SI{17}{\degree}$ (this angle is enlarged in the figure for illustration purposes). The angle $\theta_0 \approx \SI{50.4}{\degree}$ depicts the experimental acceptance angle with respect to the KATRIN beam axis.}
\label{fig:earth}
\end{figure}

The operators arise from the introduction of the Lorentz-invariance-violating four-vector $a^{\mu}$, which can be illustrated as an external vector field, as shown in fig.~\ref{fig:earth}. Lorentz invariance violation of type $a^{\mu}$ in tritium beta decay, is governed by Lagrangian contributions for each of the fermions
\begin{equation}
    L_{SME}^{a}=-\bar{\psi_{w}} a^{\mu} \gamma_{\mu} \psi_{w},
\end{equation}
where the species subscript $w$ $\in$ \{T, H, e, n\} labels the tritium,
helium, electron, and neutrino, respectively.
%, which modifies the SM Lagrangian with the term $\delta L_{\mathrm{SME}}^{a}=-\bar{\psi} a^{\mu} \gamma_{\mu} \psi$. %The Lorentz violating four-vector $a^{\mu}$ is considered constant so that energy and momentum are conserved and there is no new physics apart from CPT and Lorentz violation \textbf{Ralf}. 
%The four-vector $a^{\mu}$ can be illustrated as an external vector field, as shown in fig.~\ref{fig:earth}, which influences the endpoint $E_0$ of a non-isotropic $\beta$-decay spectrum. 
In the calculation of the $\beta$-decay spectrum, the momenta of the external particles are modified by $a^{\mu}$ and, at first order, terms $\propto a^{\mu}p_{\mu}=a^{0}p_{0}-\vec{a} \cdot \vec{p}$ appear, where $p^{\mu}$ denotes the momentum of the emitted electron. 
This term causes both a time-dependent and time-independent shift of the spectral endpoint $E_0$.
The former only occurs for non-isotropic $\beta$-sources, where $\vec{a}\cdot\vec{p}$ does not vanish. The time-dependence is caused by the rotation of the Earth in the vector field $a^{\mu}$, which leads to a temporal change of the relative direction between the electron's momentum and the vector field and hence results in a periodic change of the endpoint with sidereal frequency $\omega_{\oplus}=\frac{2\pi}{23\text{h}56\text{min}}$. The latter is caused by the isotropic part $a^0$ as well as the component $a^{z}$ along the rotation axis. 
%This modified spectrum leads to a change of the spectral endpoint $E_0$. 
%If the $\beta$-decay spectrum is observed for a non-isotropically emitted part of the electrons, $\vec{a}\cdot\vec{p}$ does not vanish and causes a modified spectral endpoint. Besides a time-independent shift, there is also a time-dependent modification of the endpoint. The latter is caused by the rotation of the Earth in the vector field $a^{\mu}$, which leads to a temporal change of the relative direction between the electron's momentum and the vector field and hence results in a periodic change of the endpoint with sidereal frequency $\omega_{\oplus}$. 
%Besides this time dependent term, there is also a time independent shift of the endpoint by the isotropic part $a^0$ as well as the component $a^{z}$ along the rotation axis. 
Usually the operator $a^{\mu}$ is expressed in spherical decomposition, where the isotropic part is represented by $(a_{\text {of}}^{(3)})_{00}$, the part along the rotation axis by $(a_{\text {of}}^{(3)})_{10}$ and the periodically time-dependent part by $(a_{\text {of}}^{(3)})_{11}$ and $(a_{\text {of}}^{(3)})_{1-1}=-(a_{\text {of}}^{(3)})_{11}^{*}$~\cite{Diaz,Lehnert:2021tbv}. 

In this work, we assume an isotropic tritium $\upbeta$-decay source and introduce anisotropy by considering a sub-set of $\upbeta$-electrons, which are emitted under a conical solid angle, defined by the experimental acceptance angle $\theta_{0}$, as illustrated in figure~\ref{fig:earth}. The endpoint $E_0$ of their energy spectrum is modified by $a^{\mu}$ in the following way:
	\begin{equation}
	\begin{aligned}
		\Delta E_{0}=& (\gamma-\beta_{rot} B \sin\xi) \frac{1}{\sqrt{4 \pi}} a_{00}^{(3)}+\sqrt{\frac{3}{4 \pi}} B \sin \chi \cos \xi a_{10}^{(3)} \\
		&+\sqrt{\frac{3}{2 \pi}} \cos \left(\omega_{\oplus} T_{\oplus}\right)\left[\left(\beta_{\text {rot }}-B \sin \xi\right) \operatorname{Im}(a_{11}^{(3)})\right.\\
		&\left. -B \cos \xi \cos \chi \operatorname{Re}(a_{11}^{(3)})\right] \\
		&+\sqrt{\frac{3}{2 \pi}} \sin \left(\omega_{\oplus} T_{\oplus}\right)\left[\left(\beta_{\text {rot }}-B \sin \xi\right) \operatorname{Re}(a_{11}^{(3)})\right.\\
		&\left.+B \cos \xi \cos \chi \operatorname{Im}(a_{11}^{(3)})\right],
	\end{aligned}
	\label{eq:endpointLV}
\end{equation}
where $\gamma$ is the Lorentz factor, $\beta_{\text{rot}}$ is the rotation velocity of Earth at the location of the experiment, $\chi$ is the colatitude of the experiment, and $\xi$ the orientation of the experimental beam-axis with respect to the local north, as shown in fig.~\ref{fig:Beamline}. The factor $B=M_{T}^{-1}(2 \pi(1-\cos \theta_0))^{-1} \pi \sqrt{E_{0}^{2}-m_{e}^{2}} \sin ^{2} \theta_0$ depends on the mass of the tritium atom $M_T$, the mass of the electron $m_e$, the endpoint without Lorentz invariance violation $E_0$, and the acceptance angle $\theta_{0}$. From this it becomes clear that an acceptance angle of less than $90^{\circ}$, and thus an anisotropy, is necessary to cause a temporal oscillation in addition to a time-independent shift.
	
In this work we use data from the first scientific run of the Karlsruhe Tritium Neutrino (KATRIN) experiment, which took place in spring 2019 and lasted about one month. The data was acquired in 361 two-hour-long scans, from each of which the spectral endpoint $E_0$ is inferred individually. This time series of $E_0$ measurements is used to search for a temporal oscillation, expressed by $\Delta E_0=A\cos(\omega_{\oplus}t-\phi)$, where the oscillation is described by an amplitude $A$ and phase $\phi$. The amplitude can directly be used to limit the Lorentz-invariance-violating parameter $(a_{\text {of}}^{(3)})_{11}$ via the relation
\begin{equation}
\begin{aligned}
    A = & \sqrt{\frac{3}{2 \pi}}\left|(a_{\text {of}}^{(3)})_{11}\right|\sqrt{B^{2} \cos ^{2} \chi \cos ^{2} \xi+\left(\beta_{\text{rot}}-B \sin \xi\right)^{2}}.
\end{aligned}
\label{eq:Avsa}
\end{equation}

\section{KATRIN experiment}
The goal of KATRIN is to measure the effective electron antineutrino mass $m_{\nu}$ with a sensitivity of \SI{0.2}{\electronvolt} at 90\% confidence level (C.L.) after about 1000 days of taking data~\cite{KATRIN:DesignReport,KATRIN:DesignReport20}. Recently, KATRIN has published the first direct sub-eV upper limit on the neutrino mass of $m_{\nu} < \SI{0.8}{\electronvolt}$ (90\% CL) based on the first two data-taking campaigns~\cite{KATRIN:2021uub}. The experiment will continue taking data for another few years.

KATRIN combines a high-luminosity windowless gaseous molecular tritium source with a high-resolution spectrometer based on the principle of magnetic adiabatic collimation with electrostatic filtering (MAC-E filter)~\cite{LOBASHEV1985305,PICARD1992345}. This combination allows to perform a precise integral measurement of the tritium $\beta$-decay spectrum in the close vicinity of the spectral endpoint $E_0\approx18.6$~keV, where the impact of the neutrino mass is maximal.

\begin{figure*}[]
	\centering
	\includegraphics[width=\textwidth]{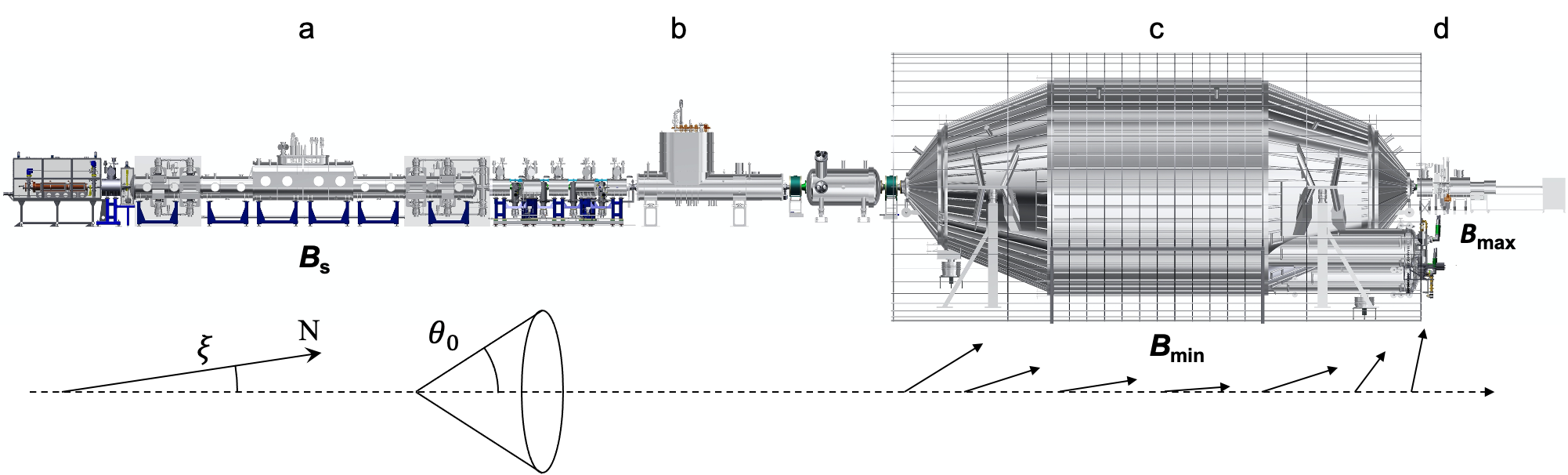}
	\caption{Main components of the KATRIN experiment: a) tritium source, b) transport and pumping sections, c) main spectrometer, d) focal plane detector. The dashed line below illustrates the KATRIN beamline. The experiment has an orientation of $\xi \approx \SI{16}{\degree}$ east to the local north N. The acceptance angle $\theta_0 = \SI{50.4}{\degree}$ defines the acceptance cone of the $\upbeta$-electrons. The black arrows below the spectrometer indicate the electron momentum (without electric field). When propagating from $B_{\text{s}}$ to $B_{\text{min}}$ the pitch angle $\theta$ (angle between the electron momentum and the magnetic field lines) is reduced, while when moving from $B_{\text{s}}$ to $B_{\text{max}}$ the pitch angle is increased. An electron starting with $\theta > \theta_0 $ will be reflected at $B_{\text{max}}$.}
	\label{fig:Beamline}
\end{figure*}

\begin{figure}[]
\centering
\includegraphics[width=\linewidth]{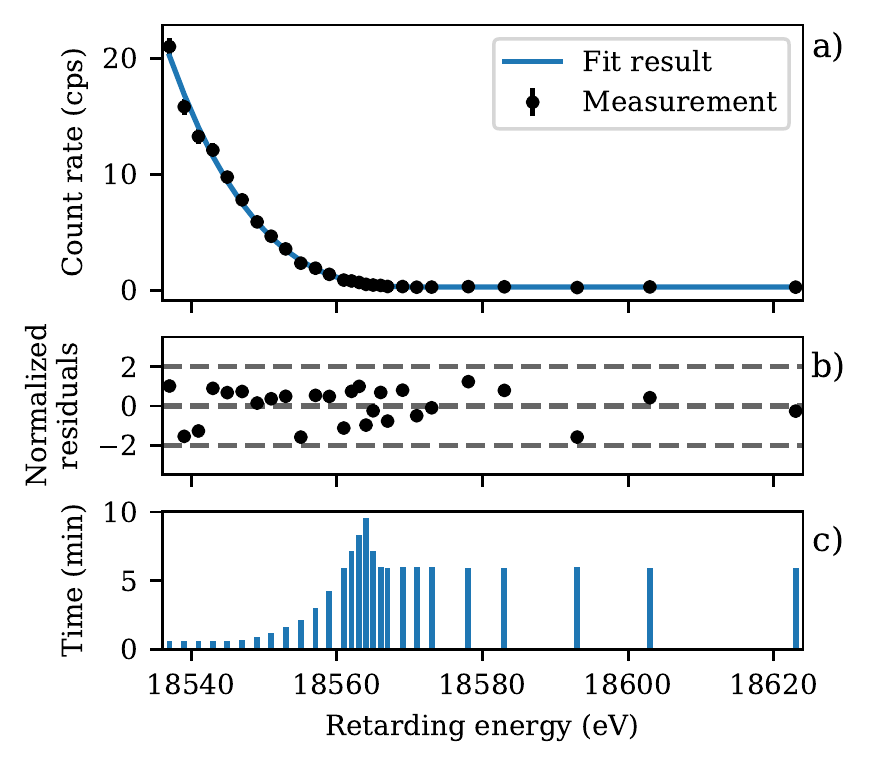}
\caption{a) Electron spectrum $R(qU)$ of a single scan where all pixels are combined (uniform fit). The spectrum of the best fit $R_{\text{calc}}(qU)$ extends to the endpoint $E_0$ and lies on top of an energy-independent background $R_{\mathrm{bg}}$. b) Residuals of $R(qU)$ relative to the 1$\sigma$-uncertainty band of the best fit model. c) Measurement-time distribution.}
\label{fig:KNM-1}
\end{figure}

Technically this is realized by a 70-m long experimental beamline, shown in fig.~\ref{fig:Beamline}, which is located at the Karlsruhe Institute of Technology (KIT) in Germany. The gaseous tritium source (a) is part of a closed tritium loop~\cite{KATRIN:TritiumLoop}, which provides up to $10^{11}$ $\beta$ decays per second in the  10 m-long, 90 mm-diameter source beam tube. The resulting $\beta$-electrons are guided by a system of superconducting magnets towards the spectrometer section~\cite{KATRIN:Magnets}. In the transport section (b), connecting the source and spectrometer, neutral and ionized tritium is removed by a differential and cryogenic pumping system~\cite{KATRIN:DPS}. The main spectrometer (c) analyzes the kinetic energy of the $\beta$-electrons with the MAC-E-filter technique. Essentially, it acts as an electrostatic filter, allowing only $\beta$-electrons with sufficient kinetic energy to overcome its precisely adjustable retarding potential $U$~\cite{KATRIN:HV}. In addition, a slowly decreasing magnetic field (from $B_{\text{max}}=\SI{4.24}{\tesla}$ to $B_{\text{min}}=\SI{0.63}{\milli \tesla}$ in the center of the main spectrometer) aligns the momenta of the isotropically created electrons. This magnetic adiabatic collimation provides a large angular acceptance ($\theta_0=\SI{50.4}{\degree}$) with a sharp cut-off energy ($\Delta E (\SI{18.6} {\kilo\electronvolt})=2.8$~eV) at the same time. By measuring the rate of transmitted electrons (charge $q=-e$) as a function of the retarding energy $qU$ the integral $\beta$-decay spectrum is obtained, as illustrated in fig.~\ref{fig:KNM-1}a. The electrons are detected by a 148-pixel silicon PIN focal-plane detector (d), installed at the exit of the spectrometer~\cite{KATRIN:FPD}.

For the presented analysis it is important that the KATRIN experiment accepts only electrons that are emitted at an an angle of less than $\SI{50.4}{\degree}$ relative to the magnetic field lines in the source. The acceptance angle $\theta_{0}=\arcsin\left(\sqrt{B_{\text{s}}/B_{\text{max}}}\right)$ is determined by the source magnetic field $B_{\text{s}}=\SI{2.52}{\tesla}$ and the maximal magnetic field in the beamline $B_{\text{max}}=\SI{4.24}{\tesla}$. These values are chosen to exclude electrons with a long path through the source and hence increased scattering probability. It is this selection of direction which makes KATRIN sensitive to the anisotropic Lorentz-invariance-violating operator $(a_{\text{of}}^{(3)})_{11}$.

This work uses data from the first high-luminosity ($2.45 \times 10^{10}$~Bq) tritium campaign, which ran from April 10 to May 13, 2019 ~\cite{KATRIN:KNM1PRL,KATRIN:KNM1Analysis}. The integral spectrum was recorded by repeatedly scanning the energy interval from $E_0-\SI{90}{\electronvolt}$ to $E_0+\SI{50}{\electronvolt}$. In each so-called scan a series of 39 non-equidistant high-voltage (HV) set points is applied to the main spectrometer. The HV set points $U$ are applied in alternating upward (up-scan) and downward (down-scan) directions to compensate for possible time-dependent drifts of the system to first order. The measurement time at each HV set point lasts between 17 and 576~s, as shown in fig.~\ref{fig:KNM-1}c. The net scan time is about 2 hours. 

The presented search for Lorentz invariance violation follows the same initial protocol as the neutrino mass analysis. To this end, we choose an analysis interval covering a region from \SI{40}{\electronvolt} below $E_0$ (22 HV set points) to \SI{50}{\electronvolt} above $E_0$ (5 HV set points). In the same manner as for the neutrino-mass analysis, we apply quality cuts to each scan and select 274 stable scans which corresponds to a total scan time of 521.7~h. The 117 selected pixels (79\% of the detector area) are combined to one effective pixel (so-called uniform fit), which leads to 274 integral spectra $R(qU_i)$ with about 7400 events each. For the neutrino mass analysis these are further combined to a single high-statistics scan, while in this work they are analyzed individually to access the temporal variation of the spectral endpoint.

\section{Endpoint fit of integral spectrum}
The theoretical spectrum $R_{\text{calc}}(qU)$ consists of the differential $\beta$ spectrum $R_{\beta}$, the experimental response function $f(E, qU)$, and a retarding-energy-independent background $R_{\text{Bg}}$:
\begin{equation}
	R_{\text {calc }}( q U)= A_{\mathrm{s}} \cdot N_{\mathrm{T}} \int R_{\beta}(E;m^2_{\nu}, E_0) \cdot f(E-q U) d E + R_{\mathrm{bg}}
\end{equation}
where $A_{\text{s}}$ is the amplitude of the signal and $N_{\text{T}}$ is the effective number of tritium atoms in the source. $E$ denotes the energy of the electron. The differential spectrum, described in detail for example in~\cite{KATRIN:KNM1Analysis, Kleesiek:2018mel}, depends on the endpoint of the tritium spectrum $E_0$ as well as the effective electron antineutrino mass $m_{\nu}^{2}=\sum_{i=1}^{3}\left|U_{e i}\right|^{2} m_{i}^{2}$, where $U$ is the PNMS matrix and $m_i$ the masses of the neutrino-mass eigenstates. In addition, it is considered that the tritium molecule can be in rotational, vibrational, and electronic excited states, which are described by means of a final-states distribution. The response function $f(E, qU)$ describes the transmission probability of an electron as a function of its energy. It includes the spectrometer resolution and energy losses due to scattering in the gaseous source.

The computed spectrum $\vec{R}_\mathrm{calc}$ is fit to the data $\vec{R}_\mathrm{data}$ by minimizing
\begin{equation}
\label{eq:chi2}
    \chi^2(\theta) = (\vec{R}_\mathrm{calc}(\vec{\eta})-\vec{R}_\mathrm{data})^{\mathsf{T}} C^{-1} (\vec{R}_\mathrm{calc}(\vec{\eta})-\vec{R}_\mathrm{data}),
\end{equation}
with respect to the free parameters $\vec{\eta}$ including the covariance matrix $C$. For the neutrino-mass analysis~\cite{KATRIN:KNM1Analysis} the combined data of all scans are analyzed with the four free fit parameters $A_{\text{s}}$, $E_0$, $R_{\text{bg}}$, and $m^2_{\nu}$. An excellent agreement of the model with the data was demonstrated in~\cite{KATRIN:KNM1PRL,KATRIN:KNM1Analysis}. In this work, in contrast, we analyze the scans separately, which leads to an individual endpoint for each scan. Moreover, we set $m^2_{\nu}$ to zero in the fits, since we assume it to be a time-independent parameter, which has no effect on the oscillation signal due to Lorentz invariance violation. An example fit can be seen in fig.~\ref{fig:KNM-1}a. 

Every fitted endpoint has an uncertainty, which is composed of a statistical and a systematic part. As we are interested in a temporal variation of the endpoint, only the statistical uncertainties and those systematic uncertainties, which can vary with time are of concern for this analysis. For example, the time-independent uncertainty on the theoretical description of the final states would have the same effect on the endpoints of all scans, and thus would not affect the oscillation signal. In contrast, the magnetic field may vary slightly from scan to scan and therefore influences the uncertainty of the fitted oscillation of the endpoint. 

Uncertainties of a statistical nature, such as the Poisson uncertainty of each data point and additional source activity and background fluctuations, are included via the covariance matrix $C$ of equation~\ref{eq:chi2}. To determine the influence of systematic uncertainties we employ the Monte Carlo propagation technique~\cite{KATRIN:KNM1Analysis}. Here, the data are fitted about $10^5$ times while varying the relevant systematic parameters (e.g. the magnetic fields and source properties) in each fit. The distribution of fitted endpoints is used to determine the best-fit value and uncertainty of the endpoint.

The statistical uncertainty of the endpoint in a single scan is $\sigma^{\text{stat}}_{E_{0}}=\SI{247}{\milli \electronvolt}$, which dominates over the total systematic uncertainty of $\sigma^{\text{syst}}_{E_{0}}=\SI{70}{\milli \electronvolt}$.  The largest effect beyond the statistical uncertainty arises from the background. Certain background sources lead to time-correlated background events and hence do not follow a Poissonian distribution~\cite{Mertens:2012vs} but instead a broader distribution. This over-dispersion of the background-rate distribution increases the statistical uncertainty and contributes \SI{69}{\milli \electronvolt} to the uncertainty budget of the endpoint in each scan. 
Uncertainties of the column density $\rho d$ (the integral of the gas density $\rho$ over the length of the source $d$), activity fluctuations during a scan, the concentration of different tritium isotopologues in the source, the source electric potential, as well as the magnetic field stability contribute with less than $\SI{10}{\milli \electronvolt}$ each to the endpoint uncertainty per scan. All uncertainties are summarized in table~\ref{tab:systematics}.

Since we are interested in a temporal oscillation of the endpoint, we verified the absence of any oscillatory behaviour of the individual slow-control and nuisance parameters. We exclude any statistically significant sinusoidal time evolution of the background rate, the source activity, and any of the systematic parameters mentioned above.

\begin{table}
	\caption{1$\sigma$ uncertainties on the endpoint $E_0$ in \SI{}{\electronvolt} for a fit to a Monte-Carlo-generated spectrum of a single scan. The values are calculated using Monte-Carlo propagation.}
	\label{tab:systematics}
	\centering
	\begin{tabular}{p{5cm}R{3cm}} \\
	\hline 
		%\hline \hline
		effect & $\sigma(E_0)$\\
		\hline \hline
		Non-Poissonian background &  0.069 \\ 
		Subrun activity fluctuation &  0.006\\
		Source electric potential &  0.005\\
		Column density $\rho d $ & 0.004\\
		Magnetic fields &  0.004\\
		Isotopologue concentration &  $<$ 0.001\\
		\hline
		\textbf{Total syst. uncertainties} &  \textbf{0.070}\\
		\hline
		\textbf{Stat. uncertainty} &  \textbf{0.247}\\
		\hline
		\textbf{Total uncertainty} &  \textbf{0.257}\\
		\hline
	\end{tabular}
\end{table}

The spectral model $R_{\text{calc}}(qU)$ assumes a constant endpoint, and thus neglects the fact that, in the case of Lorentz invariance violation, the endpoint would slightly change during the course of a 2 hour scan. We justify this approximation, by simulating several up- and down-scans, including a continuously changing endpoint, according to various assumed values of Lorentz invariance violation. We fit these simulated scans using $R_{\text{calc}}(qU)$ and demonstrate that the inferred endpoint agrees within $< 1\%$ with the theoretical true endpoint at a fixed time $t_{\text{e}}$ relative to the start of the 2-hour scan. As the spectral data points have different sensitivity to the endpoint, $t_{\text{e}}$ depends on the measurement-time distribution and the scan direction. We find $t_{\text{e}}=t_0 + \SI{87}{\minute}$ (for up scans) and $t_{\text{e}}=t_0 +\SI{52}{\minute}$ (for down scans), where $t_0$ is the start time of the scan.

\begin{figure*}[tbh]
	\centering
	\includegraphics[width=\textwidth]{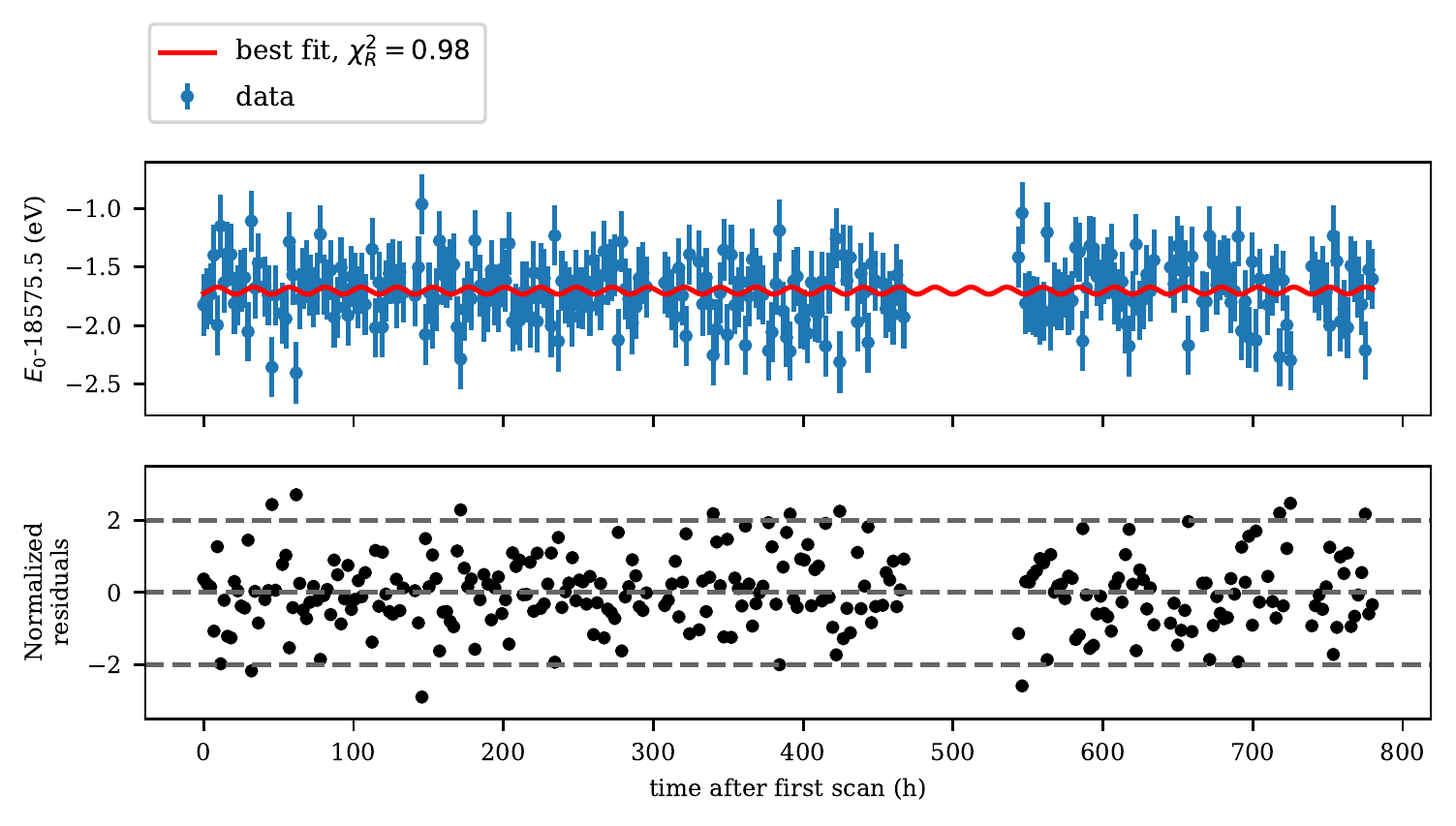}
	\caption{Illustration of the best fit of a sidereal oscillation. The fitted endpoints of the first physics campaign (blue) have been used to fit an oscillation with sidereal frequency and free amplitude and free phase. The best fit (red line) corresponds to $A_{\text{best}}\approx\SI{0.03}{\electronvolt}$ and $\phi_{\text{best}} \approx 0.78\pi$ and has a reduced $\chi^2_R=0.98$. The bottom panel displays the normalized residuals of the data fit (black).}
	\label{fig:bestfit}
\end{figure*}

\section{Search for sidereal endpoint oscillation}
The fitted endpoints with corresponding uncertainties of the individual scans, illustrated in fig.~\ref{fig:bestfit}, can be used to search for oscillations with sidereal frequency $\omega \approx 2\pi/(\SI{23}{\hour}\SI{56}{\minute})$ caused by the Lorentz-invariance-violating parameter $(a_{\text {of}}^{(3)})_{11}$. The oscillation is described by $E_0^{\text{fit}}(t_e)=D+A\cos\left(\omega t_{\text{e}}-\phi\right)$ with baseline $D$, amplitude $A$ and phase $\phi$, which are treated as free parameters. 

To set a limit on the amplitude and phase, we used a two-dimensional $\chi^2$-grid search in $(A,\phi)$. Thereby an exclusion contour is extracted from the $\chi^2$ map using a likelihood ratio as test statistic $\Lambda(A, \phi)=\frac{\mathcal{L}(A, \phi)}{\mathcal{L}_{\text {best}}}$. Applying Wilks' theorem, $\Delta \chi^{2} \equiv-2 \ln \Lambda(A, \phi)=\chi^{2}(A, \phi)-\chi^{2}_{\text {best}}$ behaves according to a $\chi^2$ distribution with two degrees of freedom~\cite{Wilks}. This means that for a confidence level of 90\%, one excludes all grid points for which $\Delta \chi^2>\chi^2_c \approx 4.61$ holds. We confirmed the correct coverage of this approach by means of Monte-Carlo simulations. For about 5000 simulations of the experiment assuming the null hypothesis $H_0$ ($A=\SI{0}{\electronvolt}$) and several alternative hypotheses $H_1$ ($A\neq \SI{0}{\electronvolt}$) we show that the $\Delta \chi^2$ distribution follows the theoretical $\chi^2$ curve and gives a critical $\chi^2$ of $\chi^2_c=\SI{4.55 \pm 0.06}{}$, which is in good agreement with Wilks' theorem. Finally, according to eq.~\ref{eq:Avsa} we translate the limit on the amplitude $A$ into a limit on $(a_{\text {of}}^{(3)})_{11}$.

%We checked the correct coverage of this approach by fluctuating the endpoints within their uncertainty for the null hypothesis $\text{H}_0$ of $A=\SI{0}{\electronvolt}$, to find the best fit and the associated $\chi^2_{\text{best}}$ as well as $\chi^2(\text{H}_0)$. We repeated this procedure about 5000 times and considered the distribution of $\Delta \chi^2=\chi^2(\text{H}_0)-\chi^2_{\text{best}}$. This follows the theoretical $\chi^2$ curve and gives a $\chi^2_c=\SI{4.55 \pm 0.06}{}$. Furthermore, we considered several alternative hypotheses $H_1$ with $A\neq \SI{0}{\electronvolt}$ and also found good agreement with Wilks' theorem. 

	% \section{Methods}
	% \subsection{Experimental setup}
	% \begin{itemize}
	%     \item KATRIN specs
	% \end{itemize}
	% \subsection{Measurement of the Tritium spectrum}
	% \begin{itemize}
	%     \item KNM-1 specs
	%     \item integrated spectrum
	% \end{itemize}
	
	% \subsection{Data Analysis}
	% \begin{itemize}
	%     \item selection of runs, pixels, etc.
	%     \item selection of fit range
	%     \item uniform fit
	% \end{itemize}
	
	% \subsection{Model of spectrum}
	% \begin{itemize}
	%     \item diff spec \& resp function
	%     \item acceptance angle
	%     \item fit parameters
	% \end{itemize}
	% \subsection{LV search}
	% \begin{itemize}
	%     \item Ralf's model for endpoint oscillation
	%     \item endpoint as fit parameter with fixed neutrino mass
	%     \item effective time (idea+studies)
	%     \item Frequentist: Wilks, \dots
	
	% \end{itemize}
	
\section{Results}
In order to prevent human-induced bias in the analysis and to determine a sensitivity of the considered data set to possible Lorentz invariance violation, the $\chi^2$ grid scan is first performed on simulated data, assuming no Lorentz invariance violation. The sensitivity to the amplitude of the sidereal oscillation is found to be $A < \SI{0.05}{\electronvolt}$ (90\% CL). This translates into a sensitivity of $\left|(a_{\text {of}}^{(3)})_{11}\right| < \SI{2.0e-6}{\giga \electronvolt}$ (90\% CL). 

\begin{figure}[]
	\centering
	\includegraphics[width=\linewidth]{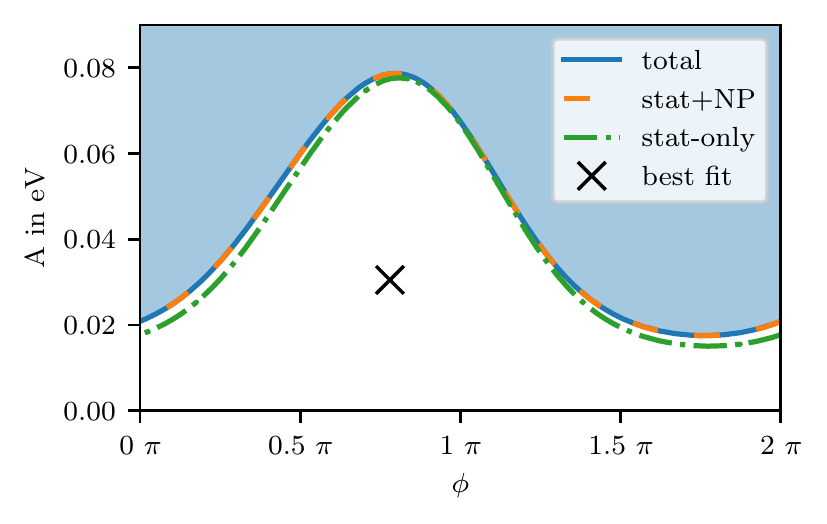}
	\caption{90\% C.L. exclusion curves for amplitude and phase of a sidereal oscillation of the endpoint in the first physics run of KATRIN. Using a $\chi^2$-grid search and Wilks' theorem the exclusion limit for the total uncertainties (blue solid line) are calculated. The green dashed-dottet line shows the exclusion limit, including only the statistical uncertainty. The orange dashed line includes an enlarged statistical uncertainty due to the non-Poissonian (NP) distribution of the background rate, as explained in the main text. The best fit is shown by the black cross. }
	\label{fig:exclusionPlot}
\end{figure}

After the investigation with simulated data, the grid search is performed with the real data, as shown in fig.~\ref{fig:exclusionPlot}. The best fit is found at $A_{\text{best}} = \SI{0.03}{\electronvolt}$ and $\phi_{\text{best}} = 0.78\pi$. The difference of the $\chi^2$ of the null hypothesis and the best fit is $\Delta \chi^2=1.86$ for two degrees of freedom, which corresponds to a p-value of 0.39 and is thus not significant at 90\% confidence level. The resulting exclusion curve, illustrated in fig.~\ref{fig:exclusionPlot}, shows that amplitudes between $A < \SI{0.08}{\electronvolt}$ (at $\phi \approx 0.78\pi$) and $A<\SI{0.02}{\electronvolt}$ (at $\phi \approx 0$) can be excluded. This null result can be transformed into the first constraint of $\left|(a_{\text {of}}^{(3)})_{11}\right| < \SI{3.7e-6}{\giga \electronvolt}$ (corresponding to $A = \SI{0.08}{\electronvolt}$) by means of eq.~\ref{eq:Avsa}.

In addition we can also constrain $(a_{\text {of}}^{(3)})_{00}$ and $(a_{\text {of}}^{(3)})_{10}$ as done in \cite{Diaz}, by searching for a time-independent shift of the Q-value of the $\beta$-decay with respect to the theoretical Q-value. In order to assess the Q-value in KATRIN the measured effective endpoint $E_{0}$ has to be corrected for the electric potential in the source $\Phi_{\text {so}}$, the workfunction of the spectrometer $\Phi_{\text{sp}}$ and the molecular recoil energy $E_{\text{rec}}$, which leads to $Q^{\text{KATRIN}}=E_{0} + E_{\text{rec}} - ( \Phi_{\text {so}}-\Phi_{\text{sp}})$, see~\cite{KATRIN:KNM1Analysis} for details. The calculated Q-value $Q^{\text{calc}}$ is given by a high-precision measurement of the mass difference of $\mathrm{^3He}$ and T~\cite{PhysRevLett.114.013003}, corrected for molecular dissociation and ionization energies~\cite{Otten:2008zz}. A significant difference between the $Q^{\text{KATRIN}}$ and $Q^{\text{calc}}$ could be interpreted as a signature of Lorentz invariance violation, caused by the parameters $(a_{\text {of}}^{(3)})_{00}$ and $(a_{\text {of}}^{(3)})_{10}$ (see eq.~\ref{eq:endpointLV}). 

In order to disentangle both parameters one needs to make use of a second experiment, for example the Mainz direct neutrino mass experiment~\cite{Kraus_2005} which is located at a different site $(\chi = \SI{40}{\degree}, \xi= \SI{-65}{\degree})$ from KATRIN. We infer both parameters by comparing $Q^{\text{calc}} = \SI{18575.72 \pm 0.07}{\electronvolt}$~\cite{PhysRevLett.114.013003} to both $Q^{\text{KATRIN}} = \SI{18575.2 \pm 0.5}{\electronvolt}$ \cite{KATRIN:KNM1PRL} and $Q^{\text{Mainz}} = \SI{18576 \pm 3}{\electronvolt}$~\cite{WEINHEIMER1993210}, which was obtained from a dedicated Q-value determination campaign of the Mainz experiment. Note that the analysis presented in \cite{Diaz} uses a different data set of the Mainz experiment. We find no significant deviation from zero for $(a_{\text {of}}^{(3)})_{00}$ and $(a_{\text {of}}^{(3)})_{10}$ and thus set an upper limit of $\left|(a_{\text {of}}^{(3)})_{00}\right| < \SI{3.0e-8}{\giga \electronvolt}$ and $\left|(a_{\text {of}}^{(3)})_{10}\right|< \SI{6.4e-4}{\giga \electronvolt}$ (90\% CL).

\section{Conclusion}
KATRIN is uniquely positioned to study oscillation-free Lorentz-invariance-violating operators that cannot be accessed by time-of-flight or neutrino-oscillation experiments. Based on the first physics run of the KATRIN experiment we were able to probe the parameter $(a_{\text {of}}^{(3)})_{11}$ by searching for a sidereal oscillation of the endpoint of the tritium $\beta$-decay spectrum. We find no oscillation and release the first upper limit on this parameter of $\left|(a_{\text {of}}^{(3)})_{11}\right|<\SI{3.7e-6}{\giga \electronvolt}$ (90\% CL). Based on the future final KATRIN data set, a sensitivity at the level of \SI{5e-7}{\giga \electronvolt} (90\% CL) could be reached. 

Besides the anisotropic Lorentz-invariance-violating parameter, the parameters $(a_{\text {of}}^{(3)})_{00}$ and $(a_{\text {of}}^{(3)})_{10}$ were investigated using the absolute endpoint measurements of both the KATRIN and the Mainz experiments. We find improved limits of $\left|(a_{\text {of}}^{(3)})_{00}\right| < \SI{3.0e-8}{\giga \electronvolt}$ and $\left|(a_{\text {of}}^{(3)})_{10}\right|< \SI{6.4e-4}{\giga \electronvolt}$ (90\% CL).

This initial study illustrates that the scientific potential of precision $\upbeta$-spectroscopy experiments, such as KATRIN, extends well beyond the neutrino-mass search to physics beyond the standard model.

\section*{Acknowledgements}
We acknowledge the support of Helmholtz Association (HGF), Ministry for Education and Research BMBF (05A20PMA, 05A20PX3, 05A20VK3), Helmholtz Alliance for Astroparticle Physics (HAP), the doctoral school KSETA at KIT, and Helmholtz Young Investigator Group (VH-NG-1055), Max Planck Research Group (MaxPlanck@TUM), and Deutsche Forschungsgemeinschaft DFG (Research Training Groups Grants No., GRK 1694 and GRK 2149, Graduate School Grant No. GSC 1085-KSETA, and SFB-1258) in Germany; Ministry of Education, Youth and Sport (CANAM-LM2015056, LTT19005) in the Czech Republic; Ministry of Science and Higher Education of the Russian Federation under contract 075-15-2020-778; and the Department of Energy through grants DE-FG02-97ER41020, DE-FG02-94ER40818, DE-SC0004036, DE-FG02-97ER41033, DE-FG02-97ER41041,  {DE-SC0011091 and DE-SC0019304 and the Federal Prime Agreement DE-AC02-05CH11231} in the United States. This project has received funding from the European Research Council (ERC) under the European Union Horizon 2020 research and innovation programme (grant agreement No. 852845). We thank the computing cluster support at the Max Planck Computing and Data Facility (MPCDF).

%\begin{acknowledgements}
%\end{acknowledgements}
\FloatBarrier
\bibliographystyle{spphys}       % APS-like style for physics
\bibliography{ref}   % name your BibTeX data base
\end{document}